\documentclass[notitlepage,english,aps,floats,onecolumn,showpacs,nofootinbib,floatfix]{revtex4-2}
\usepackage{pslatex}
\usepackage[T1]{fontenc}
\usepackage[latin1]{inputenc}
\usepackage{graphicx}
\usepackage{epsfig}
\usepackage{longtable}
\usepackage{float}
\usepackage{calc}
\usepackage{ifthen}
\usepackage{amsmath}
\usepackage{hyperref}
\usepackage{amssymb}

\usepackage{color}

{
{
{
\newcommand{\bea}{\begin{eqnarray}}
\newcommand{\eea}{\end{eqnarray}}

\newcommand{\nc}{\newcommand}
\nc{\renc}{\renewcommand}
\nc{\eqs}[2]{\mbox{Eqs.~(\ref{#1},\,\ref{#2})}}
\nc{\eq}[1]{\mbox{Eq.~(\ref{#1})}}
\nc{\figs}[2]{\mbox{Figs.~(\ref{#1},\,\ref{#2})}}
\nc{\fig}[1]{\mbox{Fig~.(\ref{#1})}}
\nc{\be}[1]{\begin{equation} \mbox{$\label{#1}$}}
\nc{\ee}{\vspace{0.1cm}\end{equation}}

\newcommand{\bean}{\begin{eqnarray*}}
\newcommand{\eean}{\end{eqnarray*}}

\def\GeV{{\rm \ GeV}}

\def\TeV{{\rm \ TeV}}

 \def\gae{\; ^{>}_{\sim} \;}

\begin{document}

\title{Higgs Inflation with Vector-Like Quark Stabilisation and the ACT spectral index} 

\author{John McDonald }
\email{j.mcdonald@lancaster.ac.uk}
\affiliation{Dept. of Physics,  
Lancaster University, Lancaster LA1 4YB, UK}

\begin{abstract} 

Recently, the Atacama Cosmology Telescope (ACT) collaboration has reported a scalar spectral index $n_s~=~0.9743~\pm~0.0034$. This is substantially larger than the classical prediction of Higgs Inflation, $n_s \approx 0.965$, which is 2.74$\sigma$ below the ACT mean value. We show that when an otherwise metastable Standard Model Higgs Inflation potential is stabilised by the addition of vector-like quark pairs and the potential is renormalised in the Jordan frame, the value of $n_s$ is generally larger than 0.965 and can explain the ACT observation. As an example, assuming the 2022 PDG direct measurement central value for the t quark mass, $m_{t} = 172.69 \GeV$, and central values for the SM inputs to the renormalisation group equations, we obtain $n_s = 0.9792 - 0.9844$ for the case of three isosinglet vector-like B quarks with mass $m_{Q}$ in the range 1-3 TeV, with the lowest value of the $n_s$ range being 1.44$\sigma$ above the ACT mean value. The model predicts primordial gravitational wave with tensor-to-scalar ratio $r = 7.87 \times 10^{-3} - 1.21 \times 10^{-2}$ for $m_{Q} =$ 1-3 TeV, which will be easily observable in forthcoming CMB experiments. Observation of vector-like quarks of mass close to 1 TeV mass combined with a large tensor-to-scalar ratio $r \sim 0.01$ would support the model.   

\end{abstract} 
 
\maketitle

\section{Introduction} Higgs Inflation \cite{bs} proposes that the inflaton is the Standard Model (SM) Higgs boson. As such, it is a strongly motivated theory based on known particle physics. To achieve the flat potential required for inflation, the model non-minimally couples the Higgs boson to the Ricci scalar \cite{bs, salopek}. When considered classically, the resulting potential is ideal for inflation and predicts a value for the scalar spectral index\footnote{This value assumes $n_{s} = 1- 2/N_{*}$, with $N_{*} = 57$ for $k_{*} = 0.05 \, {\rm Mpc}^{-1}$.} $n_s \approx 0.965$. This is in excellent agreement with the CMB result from Planck assuming the $\Lambda$CDM model, $n_{s} = 0.965 \pm 0.004$ \cite{planck}. However, the Planck $\Lambda$CDM value of $n_s$ has been challenged by a number of recent developments.  Firstly, the $\Lambda$CDM model assumed by Planck predicts a value of $H_{0}$ which is less than that directly observed from supernova and BAO observations, a discrepancy known as the Hubble tension \cite{kam}. Early time solutions to the Hubble tension, based on modifying the state of the Universe pre-recombination, generally require larger values of $n_s$. More recently, a new $\Lambda$CDM-based CMB determination of $n_s$ by the ACT collaboration \cite{act} finds $n_s = 0.9744 \pm 0.0034$, which has a mean value that is 
2.74$\sigma$ larger than the classical Higgs Inflation prediction.  

A problem for Higgs Inflation is the likely metastability of the SM Higgs potential due to quantum corrections \cite{unst1,butt,unst3}. Unless the t quark mass is close to the lower limit of its 2$\sigma$ range, the SM Higgs potential will be metastable and will turn negative at a scale $\Lambda$ which is too small to support Higgs Inflation \cite{mass}. In this case the model must be modified to increase $\Lambda$. This can be done by adding new particles to the SM to modify the quantum corrections and so increase $\Lambda$ sufficiently to recover the inflationary plateau, which requires that $\Lambda$ is greater than the scale $M_{Pl}/\sqrt{\xi}$ at which the non-minimal coupling flattens the Einstein frame potential.  

 In a previous study \cite{vlf1}, we considered stabilisation of Higgs Inflation by the addition of vector-like quarks. Vector-like quarks are anomaly-free and can have a mass term, so they are a natural way to extend the SM. In particular, adding three copies of vector-like quarks retains the structure of the SM as a theory of three fermion generations plus a single scalar multiplet.

For the case where the Higgs potential is renormalised in the Jordan frame\footnote{Prescription I renormalisation predicts a value for $n_s$ which is almost identical to the classical prediction \cite{vlf1}.}  (Prescription II renormalisation \cite{presc1,presc2}), we found  that the predictions for $n_s$ and $r$ are significantly different from those of classical Higgs Inflation, with values of $n_{s}$ in the range 0.979-0.990 and $r$ of the order of 0.01 \cite{vlf1}. In addition, the vector-like quarks are likely to have masses in the 1-10 TeV range. The values of $n_{s}$ are generally larger than the Planck CMB value,  but are compatible with the range favoured by early-time Hubble tension solutions. 

In this note we use the results of \cite{vlf1} to show that vector-quark stabilised Higgs Inflation can also agree with ACT value of $n_{s}$ to well within 2$\sigma$.

\section{A vector-like quark-stabilised Higgs Inflation model in agreement with ACT} 

As a example of a stabilised model which is in agreement with ACT, we present results for the case of three isosinglet vector-like quarks (VLQs) of mass $m_{Q}$ in the $({\bf 3}, {\bf 1}, -1/3)$ representation of ${\rm SU(3)_{c} \times SU(2)_{L} \times U(1)_{Y} }$. We refer to these as vector-like B quarks. By having the same representation as conventional b quarks, the vector-like B quarks can have Yukawa couplings to SM fields that allow them to decay, avoiding cosmological problems due to stable coloured particles \cite{cosmo}. 

The model has the Jordan frame action 
\be{e1}      S = \int d^{4} x \sqrt{-g} \left[ \left(M_{Pl}^{2} + \xi \phi^{2} \right) \frac{R}{2}   - \frac{1}{2} \partial_{\mu} \phi \partial^{\mu} \phi  - V(\phi) + \hat{ {\cal L} } \right]  ~,\ee
where $\hat{{\cal L}} $ is the Lagrangian of the SM and VLQ fields excluding the Higgs kinetic and potential terms. To analyse inflation, we transform the action to the Einstein frame via a conformal transformation $\tilde{g}_{\mu \nu} = \Omega^{2} g_{\mu\nu}$, where 
\be{e2} \Omega^{2} = \left(1 + \frac{\xi \phi^{2}}{M_{Pl}^{2}} \right)   ~.\ee 
The Einstein frame action is then 
\be{e4}  S = \int d^{4} x \sqrt{-\tilde{g}} \left[ \frac{M_{Pl}^{2}}{2} \tilde{R} -\frac{1}{2 \Omega^{2}} \left(1 + \frac{6 \xi^{2} \phi^{2} }{\Omega^{2} M_{Pl}^{2}} \right) \partial_{\mu} \phi \partial^{\mu} \phi 
 - V_{E}(\phi) + \frac{ \hat{\cal{L}}}{\Omega^{4}}  \right]   ~,\ee
where $V_{E}(\phi) = V(\phi)/\Omega^{4}$ is the Higgs potential in the Einstein frame. The conformal factor $\Omega^2$ flattens the potential once $\phi \gae M_{Pl}/\sqrt{\xi}$. 

To obtain our results, we renormalised the model in the Jordan frame (known as Prescription II renormalisation, with Prescription I being renormalisation in the Einstein frame \cite{presc1,presc2}) and then conformally transformed the renormalised potential to the Einstein frame. The Higgs Inflation models with Prescription I and Prescription II should be considered as fundamentally different Higgs Inflation models with different UV completions, which make different predictions \cite{presc1,presc2,uvframe}. 

We used the 3-loop ${\rm \overline{MS}}$ RG equations for the SM given in \cite{butt}, modified to take into account the Higgs propagator suppression $s(t)$ due to the kinetic term mixing of the physical Higgs boson with the graviton in the presence of a background $\phi$. We included the $s(t)$ factors at 1-loop given in \cite{star,corr1} and at 2-loop given in \cite{wilczek}. We supplemented the 3-loop SM RG equations with the 1-loop and leading 2-loop vector-like quark corrections given in \cite{vlqs1}. 

We assumed the 2022 PDG release direct measurement value, $m_{t} = 172.69 \pm 0.30 \GeV$ \cite{pdg1} and the mean values for other SM inputs at $\mu = m_{t}$ given in \cite{butt}:  
$g_{3} = 1.1666$, $g = 0.64779$, $g' = 0.35830$ and $\lambda_{h} = 0.12604$. We also used the relation between the t quark mass and the $\overline{{\rm MS}}$ t quark coupling at $\mu = m_{t}$ given in \cite{butt}, 
\be{r1} y_{t} = 0.93690 + 0.00556 \left(m_{t} - 173.34 \GeV\right)  ~.\ee 
The non-minimal coupling $\xi$ at $\mu = m_{t}$ was adjusted to reproduce the observed curvature power spectrum.  Instantaneous reheating was assumed, as is expected in Higgs Inflation \cite{hbb,r1,r2}, in which case the Planck pivot scale is at $N_{*} \approx 57$ \cite{vlf1}. 

\section{Results} 

The predictions for $n_s$ and $r$ as a function of $m_{Q}$ shown in Figure 1, and the corresponding numerical values of the parameters are shown in Table 1. The values of $n_{s}$ are predicted to be in the range 0.9792 to 0.9844 for $m_{Q}$ in the range 1 TeV to 3 TeV. (The present experimental lower bound on the B vector quark mass is approximately 1 TeV \cite{atlas,cms}.)    The corresponding range of $r$ is $7.87 \times 10^{-3}$ - $1.23 \times 10^{-2}$. Thus, for the assumed SM inputs at $m_{t}$, a value of $n_{s} \approx 0.980$ is obtained with $m_{Q} \approx 1-2 \TeV$. In particular, the lowest value in the range, $n_{s} = 0.9792$, corresponding to $m_{Q} = 1$ TeV, is 1.44$\sigma$ above the ACT mean value. This is a substantially better fit to ACT than the classical Higgs Inflation result, which is 2.74$\sigma$ below the ACT mean value\footnote{1.44$\sigma$ corresponds to 85.01$\%$ c.l. , whereas 2.74$\sigma$ corresponds to 99.38$\%$ c.l. .}.

Whilst the results for the VLQ-stabilised model are sensitive to the SM inputs, in particular the t quark mass and strong coupling, our results use central values and are therefore quite plausible. Our results suggest that if both ACT and VLQ-stabilised Higgs Inflation are correct, then both primordial gravitational waves and vector-like quarks will be observed by the next generation of CMB and collider experiments.

\begin{table}[htbp]
\begin{center}
\begin{tabular}{ |c|c|c|c|c| }
\hline
$m_{Q}$ 
& $n_{s}$ 
& $r$ & $\xi(m_{t})/10^{3}$ & $V_{end}\,(\GeV^4)$ 
\\
\hline
$18 {\rm \, TeV} $ 
& $0.9739$ 
& $9.20 \times 10^{-2}$ 
& $0.35$
& $1.15  \times 10^{64}$
\\
$14 {\rm \, TeV}$ 
& $0.9814$ 
& $6.94 \times 10^{-2}$ 
& $0.42$
& $1.70 \times 10^{64}$
\\
$10 {\rm \, TeV} $ 
& $0.9882$ 
& $4.06 \times 10^{-2}$ 
& $0.60$
& $1.60 \times 10^{64}$
\\
$9 {\rm \, TeV} $ 
& $0.9890$ 
& $3.48 \times 10^{-2}$ 
& $0.68$
& $1.44 \times 10^{64}$
\\
$8 {\rm \, TeV}$ 
& $0.9892$ 
& $2.97 \times 10^{-2}$ 
& $0.75$
& $1.36 \times 10^{64}$
\\
$6 {\rm \, TeV}$ 
& $0.9885$ 
& $2.12 \times 10^{-2}$ 
& $1.02$
& $1.00 \times 10^{64}$
\\
$3 {\rm \, TeV}$ 
& $0.9844$ 
& $1.23 \times 10^{-2}$ 
& $1.65$
& $6.61 \times 10^{63}$
\\
$1 {\rm \, TeV}$ 
& $0.9792$ 
& $7.87 \times 10^{-3}$ 
& $2.70$
& $4.56 \times 10^{63}$
\\
\hline
\end{tabular}
\caption{Prescription II inflation observables and parameters as a function of $m_{Q}$ for the case of three vector-like $B$ quarks.}
\end{center}
\end{table}

\begin{figure}[h]
\begin{center}
\hspace*{-0.5cm}\includegraphics[trim = -3cm 0cm 0cm 0cm, clip = true, width=0.55\textwidth, angle = -90]{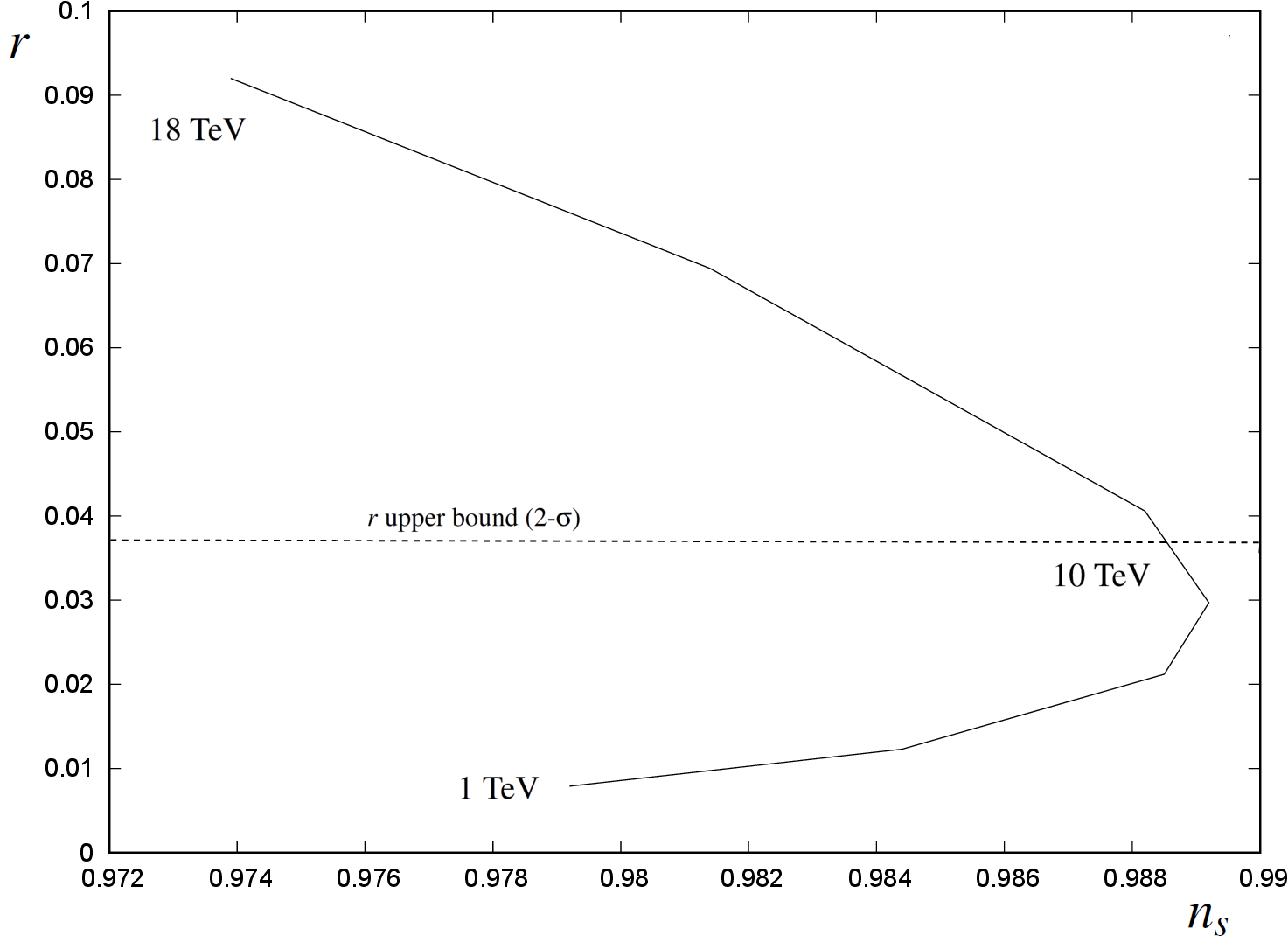}
\caption{Prescription II values of $r$ and $n_{s}$ as $m_{Q}$ varies from 1 TeV to 18 TeV, for the case of three vector-like $B$ quarks. The observational upper bound on $r$ imposes an upper bound on $m_{Q}$ of around 10 TeV. The range of values of $n_{s}$ is 0.979 to 0.989 as $m_{Q}$ increases from 1 TeV to 10 TeV. For $m_{Q} = 1$ TeV, the value of $n_{s}$ (0.9792) is 1.44$\sigma$ above the ACT mean value, with the corresponding value of $r$ being 0.0079.} 
\label{fig1}
\end{center}
\end{figure}

\section{Conclusions}    

 We have shown that prediction of Higgs Inflation for $n_{s}$ can be consistent the ACT result to well within
 2$\sigma$ when the Higgs potential is stabilised by the addition of fermions in the form of heavy vector-like quarks. For central values of the SM couplings at the renormalisation scale $\mu = m_{t}$, we find that three generations of vector-like B quarks with masses close to 1 TeV can increase the value of $n_s$ to 1.44$\sigma$ above the ACT mean value, compared to 2.74$\sigma$ below for the classical Higgs Inflation prediction, making the vector-like quark stabilised model a far better fit.  This is true for Higgs Inflation normalised in the Jordan frame (Prescription II). The model predicts a tensor-to-scalar ratio that is close to 0.01,  which will be easily observable by the next generation of CMB polarisation experiments.  In future work we aim to present a complete analysis of Higgs Inflation, its $n_s$ and $r$ predictions, and its consistency with ACT, for the unmodified SM (assuming optimistic values of the SM inputs to sufficiently stabilise the Higgs potential) and for the cases of scalar (Higgs portal) and fermionic (vector-like quark) stabilisation. 

A concern often raised is that Higgs Inflation is unitarity-violating at energies greater than $M_{Pl}/\xi$ \cite{uv0}, due to graviton-exchange Higgs scalar scattering via the non-minimal coupling. The correct statement is that perturbation theory breaks down at $M_{Pl}/\xi$, which may either indicate unitarity violation or simply that the scattering becomes a non-perturbative process that conserves unitarity. The latter is supported by a number of studies \cite{uv1,uv2,uv3}. It is also indirectly supported by the lack of any known ultraviolet completion  of Higgs Inflation ("new physics") that conserves unitarity at high energy.  In the case of strong coupling, there is no effect on Higgs Inflation, since the energy scale of inflation is the of the order of expanion rate $H \sim M_{Pl}/\xi$, which is small compared to the scale of perturbative unitarity violation in the inflationary background ($\approx \phi \gae M_{Pl}/\sqrt{\xi}$). Moreover, strong coupling does not introduce any new physics in the form of new heavy particles that might modify the Higgs potential via non-renormalisable terms.  In effect, the strong coupling interpretation decouples the ultraviolet completion energy scale of Higgs Inflation from  the energy at which perturbative unitarity breaks down, allowing the ultraviolet scale to be sufficiently large as to have a negligible effect.

There have recently been a number of other proposals to modify the Higgs Inflation prediction to be in better agreement with ACT. One is to assume that the Universe after Higgs Inflation is not dominated by non-relativistic matter or radiation, as in conventional Higgs Inflation, but is instead dominated by a form of matter with a stiff equation of state, $w > 1/3$, which increases $N_{*}$ relative to a conventional radiation-dominated era after inflation and so increases the $n_{s}$ prediction \cite{st1, st2, st3}. A different explanation proposes that threshold effects associated with a unitarity conserving ultraviolet completion modifies the Higgs potential from its SM form and introduces a mass term for the Higgs bosons at scales $\phi > M_{Pl}/\sqrt{\xi}$, which modifies the inflation predictions\footnote{Threshold effects associated with unitarity conservation were previously proposed in \cite{th1}.} \cite{wen}. In contrast, an explanation based on adding particles to the SM does not require any change to conventional reheating and maintains the link between the low-energy SM model Higgs potential and the inflaton potential,  allowing the model to be predictive.


\begin{thebibliography}{0}

\bibitem{bs} F.~L.~Bezrukov and M.~Shaposhnikov,
Phys. Lett. B \textbf{659} (2008), 703-706
doi:10.1016/j.physletb.2007.11.072
[arXiv:0710.3755 [hep-th]].


\bibitem{salopek} D.~S.~Salopek, J.~R.~Bond and J.~M.~Bardeen,
Phys. Rev. D \textbf{40} (1989), 1753
doi:10.1103/PhysRevD.40.1753

\bibitem{planck} N.~Aghanim \textit{et al.} [Planck],
Astron. Astrophys. \textbf{641} (2020), A6
[erratum: Astron. Astrophys. \textbf{652} (2021), C4]
doi:10.1051/0004-6361/201833910
[arXiv:1807.06209 [astro-ph.CO]].

\bibitem{kam}
M.~Kamionkowski and A.~G.~Riess,
Ann. Rev. Nucl. Part. Sci. \textbf{73} (2023), 153-180
doi:10.1146/annurev-nucl-111422-024107
[arXiv:2211.04492 [astro-ph.CO]].

\bibitem{act}
T.~Louis \textit{et al.} [ACT],
[arXiv:2503.14452 [astro-ph.CO]].

\bibitem{unst1} G.~Degrassi, S.~Di Vita, J.~Elias-Miro, J.~R.~Espinosa, G.~F.~Giudice, G.~Isidori and A.~Strumia,
JHEP \textbf{08} (2012), 098
doi:10.1007/JHEP08(2012)098
[arXiv:1205.6497 [hep-ph]].


\bibitem{butt} J.~Elias-Miro, J.~R.~Espinosa, G.~F.~Giudice, G.~Isidori, A.~Riotto and A.~Strumia,
Phys. Lett. B \textbf{709} (2012), 222-228
doi:10.1016/j.physletb.2012.02.013
[arXiv:1112.3022 [hep-ph]].

\bibitem{unst3}  D.~Buttazzo, G.~Degrassi, P.~P.~Giardino, G.~F.~Giudice, F.~Sala, A.~Salvio and A.~Strumia,
JHEP \textbf{12} (2013), 089
doi:10.1007/JHEP12(2013)089
[arXiv:1307.3536 [hep-ph]].


\bibitem{mass} I.~Masina and M.~Quiros,
[arXiv:2403.02461 [hep-ph]].


\bibitem{vlf1}
J.~McDonald,
JCAP \textbf{03} (2025), 055
doi:10.1088/1475-7516/2025/03/055
[arXiv:2407.02399 [hep-ph]].


\bibitem{presc1} F.~L.~Bezrukov, A.~Magnin and M.~Shaposhnikov,
Phys. Lett. B \textbf{675} (2009), 88-92
doi:10.1016/j.physletb.2009.03.035
[arXiv:0812.4950 [hep-ph]].

\bibitem{presc2} F.~Bezrukov and M.~Shaposhnikov,
JHEP \textbf{07} (2009), 089
doi:10.1088/1126-6708/2009/07/089
[arXiv:0904.1537 [hep-ph]].

\bibitem{cosmo} R.~N.~Mohapatra and S.~Nussinov,
Phys. Rev. D \textbf{57} (1998), 1940-1946
doi:10.1103/PhysRevD.57.1940
[arXiv:hep-ph/9708497 [hep-ph]].



\bibitem{uvframe} 
Y.~Hamada, H.~Kawai, Y.~Nakanishi and K.~y.~Oda,
Phys. Rev. D \textbf{95} (2017) no.10, 103524
doi:10.1103/PhysRevD.95.103524
[arXiv:1610.05885 [hep-th]].


\bibitem{star} A.~O.~Barvinsky, A.~Y.~Kamenshchik, C.~Kiefer, A.~A.~Starobinsky and C.~Steinwachs,
JCAP \textbf{12} (2009), 003
doi:10.1088/1475-7516/2009/12/003
[arXiv:0904.1698 [hep-ph]].

\bibitem{corr1}
T.~E.~Clark, B.~Liu, S.~T.~Love and T.~ter Veldhuis,
Phys. Rev. D \textbf{80} (2009), 075019
doi:10.1103/PhysRevD.80.075019
[arXiv:0906.5595 [hep-ph]].






\bibitem{wilczek}  A.~De Simone, M.~P.~Hertzberg and F.~Wilczek,
Phys. Lett. B \textbf{678} (2009), 1-8
doi:10.1016/j.physletb.2009.05.054
[arXiv:0812.4946 [hep-ph]].


\bibitem{vlqs1} S.~Gopalakrishna and A.~Velusamy,
Phys. Rev. D \textbf{99} (2019) no.11, 115020
doi:10.1103/PhysRevD.99.115020
[arXiv:1812.11303 [hep-ph]].


\bibitem{pdg1} S.~Navas \textit{et al.} [Particle Data Group],
Phys. Rev. D \textbf{110} (2024) no.3, 030001
doi:10.1103/PhysRevD.110.030001


\bibitem{hbb} F.~Bezrukov, D.~Gorbunov and M.~Shaposhnikov,
JCAP \textbf{06} (2009), 029
doi:10.1088/1475-7516/2009/06/029
[arXiv:0812.3622 [hep-ph]].


\bibitem{r1} J.~Garcia-Bellido, D.~G.~Figueroa and J.~Rubio,
Phys. Rev. D \textbf{79} (2009), 063531
doi:10.1103/PhysRevD.79.063531
[arXiv:0812.4624 [hep-ph]].


\bibitem{r2} J.~Repond and J.~Rubio,
JCAP \textbf{07} (2016), 043
doi:10.1088/1475-7516/2016/07/043
[arXiv:1604.08238 [astro-ph.CO]].




\bibitem{atlas} M.~Aaboud \textit{et al.} [ATLAS],
JHEP \textbf{05} (2019), 164
doi:10.1007/JHEP05(2019)164
[arXiv:1812.07343 [hep-ex]].

\bibitem{cms} A.~M.~Sirunyan \textit{et al.} [CMS],
Phys. Lett. B \textbf{772} (2017), 634-656
doi:10.1016/j.physletb.2017.07.022
[arXiv:1701.08328 [hep-ex]].



\bibitem{uv0} F.~Bezrukov, A.~Magnin, M.~Shaposhnikov and S.~Sibiryakov,
JHEP \textbf{01} (2011), 016
doi:10.1007/JHEP01(2011)016
[arXiv:1008.5157 [hep-ph]].



\bibitem{uv1} T.~Han and S.~Willenbrock,
Phys. Lett. B \textbf{616} (2005), 215-220
doi:10.1016/j.physletb.2005.04.040
[arXiv:hep-ph/0404182 [hep-ph]].

\bibitem{uv2} U.~Aydemir, M.~M.~Anber and J.~F.~Donoghue,
Phys. Rev. D \textbf{86} (2012), 014025
doi:10.1103/PhysRevD.86.014025
[arXiv:1203.5153 [hep-ph]].


\bibitem{uv3}
X.~Calmet and R.~Casadio,
Phys. Lett. B \textbf{734} (2014), 17-20
doi:10.1016/j.physletb.2014.05.008
[arXiv:1310.7410 [hep-ph]].


\bibitem{st1}
D.~S.~Zharov, O.~O.~Sobol and S.~I.~Vilchinskii,
[arXiv:2505.01129 [astro-ph.CO]].


\bibitem{st2}
L.~Liu, Z.~Yi and Y.~Gong,
[arXiv:2505.02407 [astro-ph.CO]].

\bibitem{st3}
M.~R.~Haque, S.~Pal and D.~Paul,
[arXiv:2505.04615 [astro-ph.CO]].


\bibitem{th1}
F.~Bezrukov, J.~Rubio and M.~Shaposhnikov,
Phys. Rev. D \textbf{92} (2015) no.8, 083512
doi:10.1103/PhysRevD.92.083512
[arXiv:1412.3811 [hep-ph]].




\bibitem{wen}
W.~Yin,
[arXiv:2505.03004 [hep-ph]].


\end{thebibliography}
\end{document}